\documentclass[10pt]{article}
\usepackage{infocomp}
\usepackage{times}
\usepackage{amsmath}
\usepackage{amssymb}
\usepackage[english]{babel}
\usepackage{graphicx}
\usepackage{hyperref}
\usepackage{enumerate}
\usepackage{caption}

\usepackage[figuresright]{rotating}
\usepackage{bm}
\usepackage{multicol}
\usepackage{titlesec}
\usepackage{latexsym}
\graphicspath{{picture/}}
\DeclareGraphicsExtensions{.pdf,.jpeg,.png,.pdf}
\usepackage{array}
\usepackage{multirow}
\usepackage{hhline}		
\usepackage{subcaption}
\usepackage{mathptmx}
\usepackage{amsfonts}
\usepackage{amsbsy}
\usepackage{adjustbox}
\usepackage{placeins}
\usepackage[utf8]{inputenc}
\usepackage{pifont}

\usepackage[inline]{enumitem}
\usepackage{tikz,pgfplots}
\usepackage{tikzscale}
\usepackage{fancyvrb}
\usepackage[para]{threeparttable}
\usepackage{MnSymbol}
\usepackage{filecontents}
\usepackage{flushend}

\usepackage{fancyhdr}
\usepackage{etoolbox}                                                        % fuer url trennung
\apptocmd{\UrlBreaks}{\do\a\do\b\do\c\do\d\do\e\do\f\do\g\do\h\do\i\do\j     % zusatz trennbuchstaben a-z
	\do\k\do\l\do\m\do\n\do\o\do\p\do\q\do\r\do\s\do\t\do\u\do\v\do\w
	\do\x\do\y\do\z}{}{}

\newcolumntype{d}[1]{D{.}{.}{#1}}
\newcolumntype{L}[1]{>{\raggedright\let\newline\\\arraybackslash\hspace{0pt}}m{#1}}
\newcolumntype{C}[1]{>{\centering\let\newline\\\arraybackslash\hspace{0pt}}m{#1}}
\newcolumntype{R}[1]{>{\raggedleft\let\newline\\\arraybackslash\hspace{0pt}}m{#1}}

\newcommand{\rfc}[2][0]{%
	\ifnum#1=0
	RFC\,#2%\xspace
	\else
	RFC\,#2~\cite{RFC#2}%\xspace
	\fi
}

\newcommand{\circled}[1]{\mbox{\ding{\number\numexpr191+#1}}}

\newcommand{\myurl}[1]{{\footnotesize\texttt{\url{#1}}}}

\address{Munich Network Management Team, Munich, GERMANY\\
		$^1$Universit\"at der Bundeswehr M\"unchen\\
		$^2$Ludwig-Maximilians-Universit\"at M\"unchen
}

\title{CAKE: An Efficient Group Key Management for Dynamic Groups}

\author{
		Peter Hillmann$^1$\\
		Marcus Kn\"upfer$^1$\\
		Tobias Guggemos$^2$\\
		Klement Streit$^1$
}

\abstract{
With rapid increase of mobile computing and wireless network linkage, the information exchange between connected systems and within groups increases heavily.
Exchanging confidential information within groups via unsecured communication channels is a high security threat.
In order to prevent third parties from accessing this data, it is essential to encrypt it.
For this purpose, the group participants need a common group key to enable encrypted broadcast messages.
But efficient key management of secured group communication is a challenging task, if participants rely on low performance hardware and small bandwidth.
For coordination and distribution, we present the modular group key management procedure CAKE that is centrally organized and meets strict security requirements.
The lightweight G-IKEv2 protocol in combination with the key exchange concept of CAKE leads to an efficiently integrated solution.
The hybrid approach combines the advantages of the existing protocols with the objective to reduce the computation and communication effort. 
It is shown that the procedure is more suitable for changing MANET groups than the existing ones.
Moreover, the exchanged group key can be used for any services which provides a wide range of applications.
}


\begin{document}

\maketitle

\newpage

%% main text
%\vspace{-0.3cm}
\section{Introduction}\label{sec:introduction}
%\todo{Marcus: - komplette erste Seite inklusive Bild für gesicherte Gruppenkommunikation für IoT oder mobiler Knoten - z. B. artistische Drohnenflüge in Gruppen}\\

%- Motivation: Warum wird CAKE  in Verbindung mit einer gesicherten und leichtgewichtigen Kommunikation gebraucht?

In today's interconnected world, the wireless network linkage is growing rapidly.
More and more devices are connected, especially in mobile and resource-constrained networks.
Coming from fixed line networks, unicast communication have been predominant so far.
During the last decade, we have witnessed the rise of new low power network technologies, such as ZigBee, IEET\,802.15.4, Bluetooth, LoRa, SigFox, 4/5\,G, just to name a few.
Most of these have in common, that the bandwidth is limited and needs to be managed wisely, while the amount of information explodes with new scenarios in the context of IoT or Smart X. 
In order to meet the requirements in constrained environments, group communication is becoming more important for information exchange, as the benefits are reduced network overhead, computation power, and energy consumption.
Efficiency is achieved by transmitting data packets only once, simultaneously to all group members.
However, when it comes to security and privacy multicast lacks efficiency.
%
%A security risk arises, when confidential information is exchanged via unsecured communication channels.
Major challenges arise from the necessary management of the multicast group (in the following referred as communication group), which needs to be secured as well as the actual communication within the group.

In order to exchange confidential data within a group in a secure manner there exist methods to manage group keys, so called \textit{Group Key Management Protocols }(GKMP).
These protocols enable the secure access as well as the secure and efficient exchange of relevant information, such as group keys. 
Group keys are used for the encrypted communication within a group.
Typically, all Group Members (GM) posses a symmetric key which is used to encrypt and decrypt the exchanged data.
After all, it does not matter which service is using the GKMP infrastructure as underlying security technology. 

\begin{figure}[hbt]
	%{\centering
	\centering
	\captionsetup{justification=centering}
	\includegraphics[width=0.48 \textwidth]{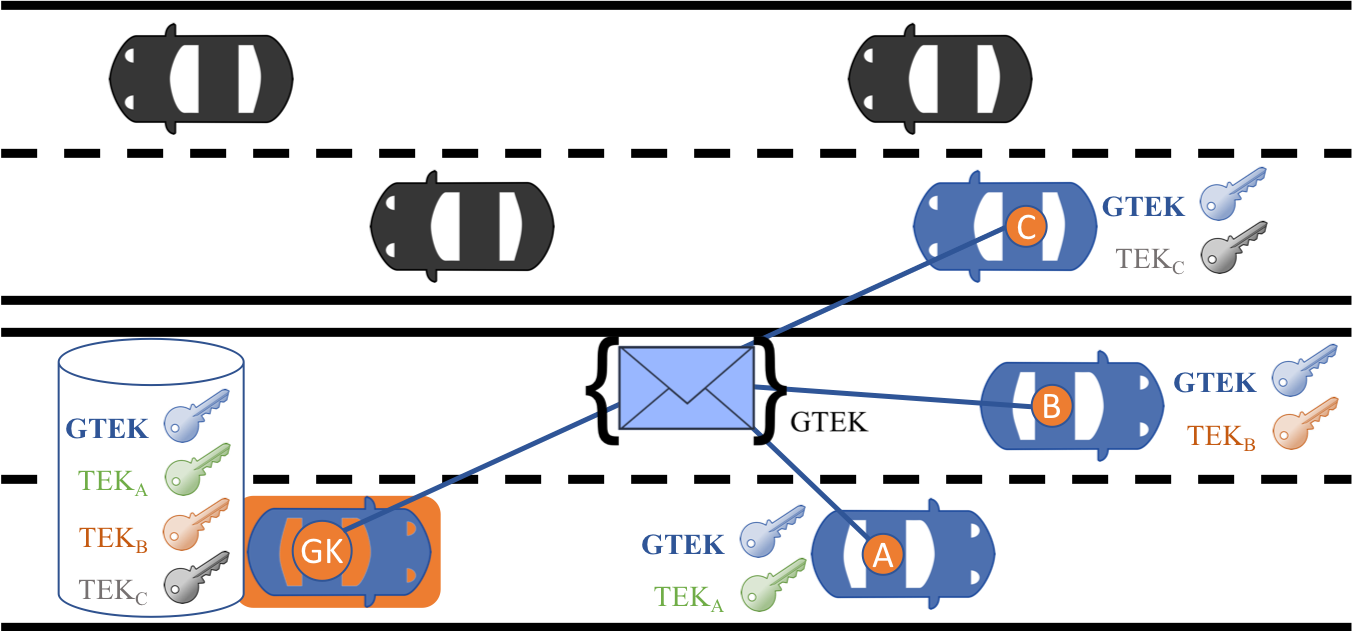}
	\caption{Secure and Efficient Group Communication in a Vehicular AdHoc Network (VANET)}
	\label{fig:groupcomm}
	%}
	%\vspace*{-0.2cm}
\end{figure}

Different areas of application do have distinct requirements to be covered by a group key management concept.
Consequently, no existing concept could be applied universally to all areas.
Depending on the dynamics of the group composition and changes, the calculation of cryptographic keys for the group communication, and the numbers of transmitted messages it could be demanding for the GMs concerning computation power, network bandwidth and energy consumption.
As a consequence, existing solutions for secured group communication do not fulfill the requirements for mid-sized and large networks in resource-constrained environments.
Exemplary use cases as in Figure~\ref{fig:groupcomm}, where an efficient and secure group communication is needed, are \textit{Internet of Things} (IoT) networks, saftey-critical system networks, \textit{Wireless Sensor Networks} (WSN), and  \textit{Mobile AdHoc Networks} (MANET) for public authority communication.

%In this paper, a hybrid concept is introduced to improve the efficiency and security of group communication in constrained enivronments.
%This is achieved by using the currently proposed \textit{Group-IKEv2} (G-IKEv2)~\cite{yeung-g-ikev2}, which is based on \textit{Internet Key Exchange v2} (IKEv2)~\rfc[1]{7296} as communication protocol on the one hand and the group key management concept \textit{Central Authorized Key Exchange} (CAKE)~\cite{Hillmann2017} on the other hand.

The paper at hand proposes a mechanism to optimize the distribution and updating procedure of keys and thus aims on improving the general security properties in group communication in constrained environments.
The contribution of this paper is the combination of an efficient networking protocol for group key management with a tree-based key update mechanism and a cryptographic key distribution scheme.
The centralized \textit{Group-IKEv2} (G-IKEv2~\cite{RFC7296,yeung-g-ikev2}) protocol is used for communication and initial key exchange with a key server. % taking care of the actual group management,
Thereon, the mechanism called \textit{Central Authorized Key Extension} (CAKE~\cite{Hillmann2017}) is used to realize encrypted group communication.
It utilizes the idea of \textit{Secure Lock}~\cite{chiou} for compressing secured information into a single cipher and enhances its performance by the combination with a logical key hierarchy (LKH~\cite{sakamoto}) and new concepts for managing keys.

%- 1 Bild auf der ersten Seite, am Besten 2. Spalte ganz unten
% picture mandatory on the first page
%\vspace*{-0.1cm}

%- Wofür kann man es überall einsetzen? Sicherheitskritischer Systeme, IoT, Sensornetze, Behorden Kommunikation
%- Definition der Begrifflichkeiten

%- ggf. Kurze Info zum Aufbau des Papers, sofern Platz
%Section 5 gilt als Eigenname, das es genau eine spezifische Section referenziert... ---> Grossschreibung
The remainder of this paper is structured as follows.
In Section II a scenario illustrating the need of this hybrid approach as well as the requirements is given.
Section~\ref{sec:sota} introduces related work on secure group communication and evaluates security as well as efficiency regarding the given scenario with a special focus on the \textit{Logical Key Hierarchy}~(LKH).
%Section III introduces related work on secure group communication and evaluates security and efficiency regarding the given scenario.
Afterwards, Section~\ref{sec:concept} describes the concept of our hybrid system combining the light-weight G-IKEv2 protocol and \textit{Central Authorized Key Extension}, called CAKE.
%The evaluation of this concept is given in Section V, before Section VI summarizes and concludes this paper. 
Finally, Section~\ref{sec:evaluation} evaluates the findings and compares with the well-established LKH, before Section~\ref{sec:conclusion} summarizes and concludes this paper.

%On basis of this, Section~\ref{sec:concept} describes the concept of this work combining the lightweight G-IKEv2 and \textit{Central Authorized Key Extension}.

\section{Scenario and Requirements}\label{sec:scenario}

This work aims at concepting a highly efficient, but secure group key management scheme to facilitate secure group communication.
The motivation is mainly found by the manifold use of constrained devices in a wide range of applications such as civil, industrial or military use cases.
A prominent example for a civil application is car-to-car communication (see Figure~\ref{fig:groupcomm}), mainly revealing highly dynamic group formations and wireless network limitations.
In contrast, in home automation scenarios (think of smoke detectors on battery) or military applications such as head-mounted units limitations in terms of availability of power, main memoy and storage, CPU and network datarates prevail.

Encrypted communication among a set of more than two group members is common to all scenarios.
Thus, it seems desirable to share one cryptographic key among the group members in order to encrypt message transfers.
Unfortunately, the management of such a key becomes costly quickly due to the dynamics withing a group (i.\,e. members joining or leaving the group rather frequently).
Changing the cryptographic material upon every single group management action seems unavoidable, which motivates working towards other than naive approaches.
Analyzing diverse application areas leads to a set of requirements that can be organized into mostly two categories~-- security requirements and scenario-driven non-functional requirements.

%\todo[inline]{Peter: Vorschlag: Bild von Marcus der Introduction hier einfügen und anhand dessen die Anforderungen erläutern. @Marcus: Bild für eine Anwendung am Ende der Introduction: fliegende Dronen einer Lichtshow - zivile Nutzung.}

\subsection*{Security requirements}

\begin{description}

  \item[Forward Secrecy:] Whenever a group member leaves the group or is expelled, the member in question must not be able to have access to a valid group key.\\
  
  \item[Backward Secrecy:] Whenever a new group member joins a group, the member in question must not be able to have access to a formerly valid group key before joining.
  
  \item[Key Independence:] Having access to one key must not yield the possibility to deduce another member's key.

%\todo[inline]{ - Antwort Peter: Kollisionsfreiheit: Eine unerlaubte Vorgehensweise einer Teilmenge von Teilnehmern führt nicht zum Schaden eines einzelnen Gruppenteilnehmers. Schlüsselunabhängigkeit/Folgenlosigkeit:Die Kenntnis eines Schlüssels ermöglicht keine Schlussfolgerung auf weitere Schlüssel.}

  \item[Collision Free:] Additionally, specific to group communication, there must not be a subset of group members that can deduce another member's key(s) by combining their knowledge.
  %Any cryptographic material shall be free of duplicates\,/\,collisions with a probability close to~100\,\%.
  
  \item[Minimal Trust:] A certain trust relationship will be found mandatory, but it shall be subject to minimize.
  
  \item[``CIA:'']-- Confidentiality, Integrity and Authenticity must be granted any time, also implying the avoidance of man-in-the-middle attacks or data injections.

\end{description}

\subsection*{Scenario-driven non-functional requirements}

\begin{description}

  \item[Low Datarates:] The amount of transmitted data shall be minimal in order to facilitate network limited applications.

  \item[No 1-to-n Effect:] Limited impact of a single membership change on all the other group members is mandatory, meaning not suffering from the 1-affects-n phenomenon, if a single membership change in the group affects all the other group members.
  %This happens typically when a single membership change requires that all group members commit to a new TEK - Bei einigen Protokollen ist dieser Effekt extrem Stark... z.B. bei MIKE bei einem Austritt muss ein Client ggf. 10 neue Schlüssel erzeugen und berechnen

  \item[Minimal Delay:] The delay imposed by the use of both management actions and cryptographic operations must be minimal.  
  
  \item[Minimal amount of key changes and exchanges:] The amount of management actions such as exchanging (new) keys shall be limited to a necessary minimum (not implying anything about the total amount of keys in general).

  \item[Low calculation complexity:] Especially in scenarios exposing CPU limitations, a low complexity of cryptographic calculations is vital, while keeping up the maximally possible security level at the same time.
	
\item[Compatibility:] Clients not capable or not willing to support fancy optimizations should not be excluded from the communication. This is why a potential fall-back to simple (and standardized) mechanism should be supported. %\\\todo[inline]{Vielleicht interessant, vor allem im Zuge der Entscheidung zu G-IKEv? - Antwort Peter: ja, das ist gut hinsichtlich Fehlerresistenz. Gemäß RFC Policy. Send strict. Recieve maximum tolerant}
    
\end{description}

%\todo[inline]{Peter Gruppenoperationen:
\subsection*{Scenario-driven functional requirements}
Based on the scenario, this yield in the following requirements for the group operations to be supported while complying the security requirements:
\begin{description}
	\item[Join:] One or more participants accede to an existing group. (Backward Secrecy)
	\item[Leave:] One or more group members quit the group membership. (Forward Secrecy)
	\item[Re-Keying:] Updateing the group key using an efficient procedure. (Prevent statistical analysis)
	\item[Merge:] A common key can be efficiently provided to several groups by re-keying. (Backward Secrecy)
	\item[Split:] A group is divided into several subgroups. (Forward Secrecy)
\end{description}

%}
%\item[Compatibility:] Clients not capable of supporting CAKE should not be excluded from the communication. Thus, a potential fallback to standardized mechanisms should be supported.

\section{Related Work}\label{sec:sota}
%\todo{Peter + Tobias: 1,5 Spalten}\\

Secure group communication is an extensively studied area and resulted in a couple of standardization activities (most recently a new standardization group for group key distribution was formed within the IETF\footnote{\url{https://datatracker.ietf.org/wg/mls} - Started in February 2018}).

Rafaeli et~al.~\cite{Rafaeli:2003} survey a set of approaches for secure group key distribution~(GKD).
According to their analysis, there are three different types of GKDs: centralized, decentralized and distributed GKD protocols.
Most of the protocols considered are rather \textit{Cryptographic Key Schemes~(CKS)} than networking protocols, but some of them are included in \textit{Group Key Management Protocols}.
The paper at hand offers the integration of an optimized \textit{Cryptographic Key Schemes} into a centralized management protocol.
Thus, the following section is divided in \textit{Group Key Management Protocols} for communication and \textit{Cryptographic Key Schemes} to manage the group key.
To our knowledge, none of the approaches provides an efficient and integrated solution, especially with focus on low resource requirements.
This is one of the reasons why the \textit{Internet Engineering Task Force} (IETF) started a standardization process for group key distribution in February 2018 \cite{Sullivan2018}.

%\subsection{Group Key Management Architecture}
\subsection{Group Key Management Protocols}\label{GKMP}
A high-level definition of \textit{Group Key Management Protocols}~(GKMP) and their corresponding architecture is given by the IETF standard body in \rfc[1]{2093} and \rfc[1]{2094}.
The development of actual GKMPs builds on top of these specifications and usually goes hand in hand with the development of a peer-to-peer key exchange protocol and its corresponding architecture.
The \textit{Internet Security Association and Key Management Protocol}~(ISAKMP, \rfc[1]{2408}), and \textit{Group Domain of Interpretation}~(GDOI, \rfc[1]{6407}) have been the first instantiations.
The requirements and design of these protocols were derived from multicast architectures of network vendors.
Both, peer-to-peer key exchange and Group Key Management were revised for the sake of stronger security properties and better performance, resulting in \textit{Internet Key Exchange~v2} (IKEv2, \rfc[1]{7296}) and the currently proposed G-IKEv2~\cite{yeung-g-ikev2} for groups.
As G-IKEv2 offers a reduced networking overhead and includes a structure for distributing hierarchical keying information, we build the design of our solution on top of G-IKEv2 and the latest architecture from \rfc[1]{4046}.

%As G-IKEv2 offers a minimal networking overhead~\cite{gentschendFelde.2017} and includes a structure for distributing hierarchical keying information, we build the design of our solution on top of G-IKEv2 and the latest architecture from \rfc[1]{4046}.

\subsection{Cryptographic Key Schemes}\label{CKS}
Centralized Cryptographic Key Schemes (CKS) comprise a central control authority to manage the group key and to coordinate the cryptographic procedures, often based on a GKMP.
%In the decentralized approach, several instances share the management of the keys.
%In contrast, all members of decentralized techniques are equally and are involved in generation of the group key.
In contrast, decentralized techniques share the management of the keys between several instances~\cite{Challal,rafaeli}.
Thereby, the generation and distribution of group keys is realized by cooperative instances, which are typical hierarchically ordered.
In addition, distributed key agreement procedures delegate the key generation process to not only an individual group member, but to a group of members.

One example is the \textit{Group-Diffie-Hellman} Key Exchange~\cite{Steiner.2000}, but others exist~\cite{Rafaeli:2003}.
All members of a group are organized in a virtual topology, typically into a ring, hierarchies on basis of trees, or just unstructured.
%Consequently, each member is involved on the key generation process.
In all these schemes, every member of a group shares a common \textit{Transport-Encryption-Key}~(TEK).

Another approach is dividing groups into subgroup with individual TEKs.
A master within every subgroup takes care of the communication and keys, which allows avoiding \mbox{1-to-n} effects while re-keying~\cite{Challal}.
The downside is requiring repetitive conversions of encrypted messages between the subgroups.
Within the subgroups, these approaches use key management techniques of the three shown categories why out of scope of this work.

%Each main category can be further divided into subcategories.

Despite their structured nature, centralized CKS can further be categorized into one of the three subcategories:
\begin{itemize}
\item Pairwise keys: Transmission of the group key by the central instance via individual subscriber communication
\item Broadcast secret: Transmission of the group key via broadcast instead of individual secured connections
\item Hierarchical structure: Coordination of participants in a tree structure with corresponding cryptographic subkeys
\end{itemize}

%
%Nachteile verteiltes Schlüsselmanagement:
%- mehrere Instanzen sich die Verwaltung der Schlüssel teilen, sodass ein größerer Aufwand für die Schlüsselgenerierung
%- Jeder Teilnehmer muss Berechnungen durchführen: 1 zu n Effekt, was auch Auswirkungen auf den Delay hat, da man auf alle Teilnehmer warten muss.
%-  Engpässe bei der Verteilung von Schlüsseln (TEK, ...) an die Gruppenmitglieder vermieden und der Ausfall einer Instanz verursacht nicht den Ausfall des gesamten Syste

The first and most widely recognized CKS ever is defined in the GKMP, which belongs to the category of the pairwise keys.
The central server shares an individual secret key with each group member, which is called the \textit{Key-Encryption-Key}~(KEK).
For a common TEK of a group, the server generates these.
Subsequently, the server sends the group key to each participant individually encrypted using the KEK.
Upon change of the group constellation, the entire group is re-created, leading to high management and communication overheads.

An example for the broadcast secret is the \textit{Secure Lock}~(SL)~\cite{chiou,antosh} that enables the creation of a group or a re-keying action using a single broadcast message.
The SL scheme is based on the \textit{Chinese Remainder Theorem}~(CRT)~\cite{zheng,Xu2012}, which uses the properties of congruence to encrypt.
However, the reduction of communication overhead is obtained by more complex calculations compared to GKMP so that this approach only renders feasible in special scenarios.

A compromise are schemes building on hierarchical structure.
A well-known approach is \textit{Logical-Key-Hierarchy}~(LKH)~\cite{sakamoto,liu}, which is integrated into GDOI and G-IKEv2.
The KEK's and the group participants are maintained in a binary tree.
Each node in the tree represents a KEK that is known to the underlying nodes.
Maintaining the associated keys of the tree structure increases the management effort, especially the calculation and distribution of internal keys.
This approach offers a moderate advantage only in case of repetitive leavings of group members.
Since this operation does not take place in every secured group, this is unnecessary effort.

Focusing on the motivation for this paper, a centralized scheme with common TEK renders mandatory, especially in order to control and authorize individual members of a group.
In this paper, a combination of the advantages of GKMP, SL and LKH as CAKE~\cite{Hillmann20172} with an integration into G-IKEv2 is proposed, allowing for efficient key management.
%In the evaluation in Section~\ref{sec:evaluation} we will compare our results against LKH, which fits best into our approach in defining integration standardized protocols.

%\vspace*{-0.3cm}
\section{Concept}\label{sec:concept}
%1. Beschreibung der Krypto pro Operation
%2. Beschreibung der Kommunikation pro Operation

% %To meet the security and communication requirements, we propose the following, integrated concept.
%For a highly efficient and encrypted group communication, we propose the following, integrated concept.
%The developed approach \textit{Central Authorized Key Exchange (CAKE)} uses individual components of the aforementioned protocols in Section~\ref{sec:sota}.
% %These are effectively combined to a hybrid system, which is based on~\cite{gentschendFelde.2017,Hillmann2017}
% These are effectively combined to a hybrid system, which is based on: \textit{G-IKEv2}~\cite{gentschendFelde.2017} for communication and \mbox{\textit{\underline{C}entral \underline{A}uthorized \underline{K}ey \underline{E}xtension}~\cite{Hillmann2017}} for group key management.

Targeting highly efficient and encrypted group communication, this paper proposes the combination of lightweight \mbox{\textit{G-IKEv2}}~(\cite{yeung-g-ikev2}) for the key exchange and \textit{\underline{C}entral \underline{A}uthorized \underline{K}ey \underline{E}xtension}~(CAKE)~\cite{Hillmann2017} for the group key management.
CAKE's key management is centrally organized and requires a trustworthy \textit{Group Controller}~(GC).
The GC is responsible for the generation, administration and distribution of the keys and thus requires more computational power than any other (lightweight) group member.

The remainder of this section is organized into subsections inspired by group management operations and patterns:
\begin{enumerate}[label=(\Alph*)]
	\item Client-Server communication based on G-IKEv2
	\item Member Registration on the GC
	\item Group and Group Key Creation
	\item Re-Key of the group
	\item Join of member(s) to a secured group
	\item Leave\,/\,Exclude of member(s) from a secured group
	\item Tree Management and Key Addressing
	\item Merging and splitting groups
\end{enumerate}
%In the end, a \textit{Protocol Overview} is given to summarize the concept.

\subsection{Client-Server communication based on G-IKEv2}
G-IKEv2~\cite{yeung-g-ikev2} is used to secure the transmission of cryptographic material for CAKE as it has already proven suitable for constrained devices~\cite{gentschendFelde.2017}.
G-IKEv2 already supports the establishment of a confidential and authenticated \mbox{1-to-1} channel between a client and the GC.
It also offers the distribution of \textit{Group Transmission Encryption Key}s~(GTEK) and \textit{Group Key Encryption Key}s~(GKEK) and thus only requires additional support for CAKE.
To communicate securely in a group, every group member has to possess a GTEK used for the communication in the group and a GKEK used to distribute the GTEKs securely.
Figure~\ref{fig:gikev2-exchange} gives an overview about G-IKEv2:

\begin{figure*}[]{%
    \centering
    \includegraphics[width=.92\textwidth]{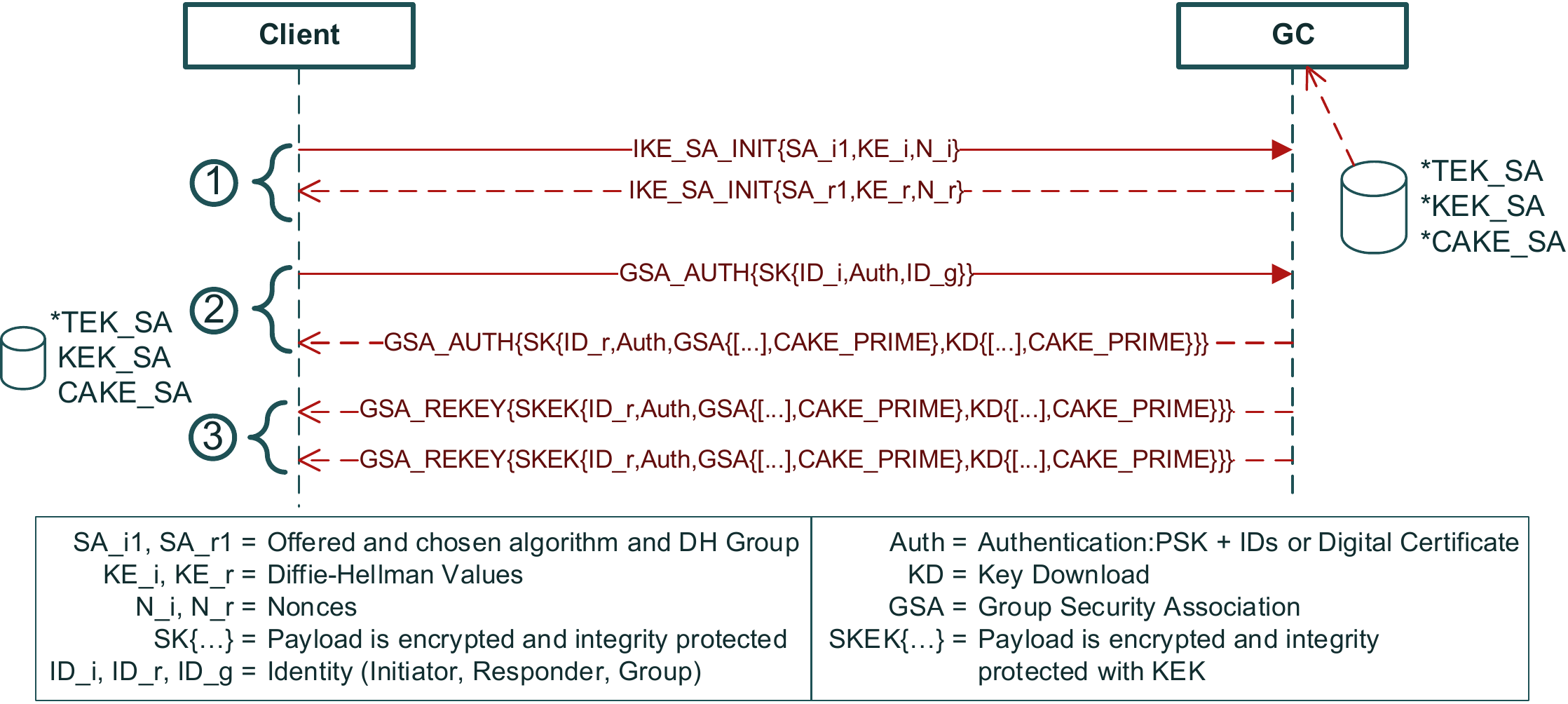}
    %\vspace*{-0.1cm}
    \caption{G-IKEv2 exchange with CAKE features}%
    \label{fig:gikev2-exchange}%
    %\vspace*{-0.2cm}
}\end{figure*}

\begin{enumerate}
  \item \textbf{Key Exchange}: A G-IKEv2 key exchange can be divided into two phases:
    \begin{enumerate}
      \item Establishing an \textit{Initial Security Association}~(IKE\_SA\_INIT): The first two messages from the client to the GC and back establish a \textit{Security Association}~(SA) and thus a secure channel between the client and the server (Phase~\circled{1}: Initialization).
      \item Exchanging keys~(GSA\_AUTH): Given the secured communication path, the client identifies and authenticates itself and in turn receives transport and key encryption keys (GTEK and GKEK) from the server. The \textit{Group Security Association~(GSA) Policy} includes the security parameters (algorithms, lifetime, etc.), while the actual keys are transported within the \textit{Key Download~(KD) Payload} (Phase~\circled{2}: Group Lifetime).
    \end{enumerate}
  \item \textbf{Re-Keying}~(GSA\_REKEY): Whenever a GTEK or GKEK loses validity (e.\,g. being outdated), a re-keying action is triggered by the server~(GSA\_REKEY), which is close to equal to the GSA\_AUTH phase (Phase~\circled{3}: Group Key Refresh).
\end{enumerate}

\subsection{Member Registration on the GC}
Each participant $P_{i}$ registers with CAKE by negotiating an individual key pair ($Key_{i}$) with the GC during an IKE\_SA\_INIT exchange~(\circled{1}).
The initial exchange is done with a Diffie-Hellman key exchange, which by design lacks authenticity.
A second message GSA\_AUTH~(\circled{2}) is used to authenticate both, the client and the GC.
Note, that the GSA\_AUTH can be used to directly join a group as part of the registration process (see Section~\ref{sec:concept-join}).

\subsection{Group and Group Key Creation}
On request, the GC randomly generates a GTEK and GKEK. % with cryptographic strong methods. % Referenz?
According to G-IKEv2, the GC manages cryptographic material and algorithms for every group.
They are stored in the \textit{TEK\_SA} and \textit{KEK\_SA} databases (see Figure~\ref{fig:gikev2-exchange}).

The GC may decide to create a new group with the new group key and members already registered and authenticated by building a GSA\_REKEY payload as follows:
\begin{enumerate}
	\item The GC constructs a CRT congruency in analogy to the SL scheme, so-called \textit{Lock~MX}.
	Therefore, it uses the individual $m_{i}$ and $Key_{i}$ from all participants of the specified group to calculate the Lock MX to encrypt the GKEK (see CRT calculation~\cite{zheng,Xu2012}).
	\item The GTEK is encrypted with the GKEK. 
				For the sake of efficiency and security~\cite{Matt}, $XOR$-operations are used for bitwise encryption of the new key tuple with a hashed GKEK.
				However, any encryption method specified by G-IKEv2 is supported.
	\item The keys are embedded into a \texttt{CAKE\_PRIME} GSA Policy (including the new \texttt{KEK\_MANAGEMENT\_ALGORITHM} called \texttt{CAKE}) and a \texttt{CAKE\_PRIME} KD payload.
                They are distributed using a single GSA\_REKEY broadcast message.
\end{enumerate}

A participant $P_{i}$ can only ``open'' the Lock MX, if she possesses a value $m_{i}$ that was included during the creation of the lock.
In consequence, only intended recipients (i.\,e. group members) are able to read the GKEK and GTEK by solving the CRT.
% So the valid recipients of the message receive the GKEK as a result of the CRT congruency and ensuing the encrypted GTEK.
% This means that all group members are familiar with the uniform group keys GTEK and GKEK.

\subsection{Re-Key of the group}\label{sec:concept-rekeying}\label{cake:rekeying}
In case the $GTEK$ needs to be renewed, a re-keying action is carried out.
%Please note that re-keying can also be used to grant forward and backward secrecy.
The GC generates the keys $GKEK_{new}$ and $GTEK_{new}$, which will be encrypted using the $GKEK_{current}$, embedded into the KD and broadcasted with a GSA\_REKEY message.
In order to grant forward and backward secrecy, a re-keying action is also carried out every time a member joins or leaves the group. 
%Both action use the \textit{GSA\_REKEY} message as well.

\subsection{Join of new member(s) to a secured group} \label{sec:concept-join}
If a new participant $P_{i+1}$ wishes to join the group, she sends a GSA\_AUTH request including the group ID $Id_g$ she wishes to join.
The GC authenticates $P_{i+1}$ and generates an inhomogeneous prime number $m_{i+1}$ for a CRT congruency for $P_{i+1}$.
Additionally, a new GSA policy and KD payload called \texttt{CAKE\_PRIME} is added, holding $m_{i+1}$.
The use of CAKE is communicated with a new \texttt{KEK\_MANAGEMENT\_ALGORITHM} called \texttt{CAKE} within the GSA Policy (see Section~4.5.1.1 in~\cite{yeung-g-ikev2}). 
The GC also generates $GKEK_{new}$ and $GTEK_{new}$ and embeds the information and keys into an GSA\_AUTH sent to the new group member via unicast.

Additionally, a re-key is triggered for any of the former group members.
The re-key includes the KD $(GKEK_{new}, GTEK_{new})$ encrypted with the $GKEK_{Current}$.
Switching from $GTEK_{old}$ to $GTEK_{new}$ enables the enlarged group (former group plus joined members) to communicate securely.
As long as the $GTEK_{old}$ and $GKEK_{old}$ are still secure the $GTEK_{new}$ and $GKEK_{new}$ should be generated by hashing the $old$ once.
So the keys do not need to be distributed over the network. Only a tiny information message is necessary.

A mass entry of more than one new participant is equivalent to the process as described before, whereas the $GTEK_{new}$ is send to every new member individually. %only one rekeying need to be performed.
Alternatively, the new participants can be combined together via a CRT to transmit the $GKEK_{new}$ so that only one message is necessary instead of multiple individual ones.
Unfortunately, the latter is only possible if the joining clients are already authenticated.
In both cases, two GSA\_REKEY messages are broadcasted, one holding the Lock MX for the new clients and one with $(GKEK_{new}, GTEK_{new})$ encrypted with the $GKEK_{Current}$ for the former group members.
Thus, an arbitrary number of new members joining a group requires a constant number of messages and thus scales efficiently with the amount of new members.

\subsection{Leave\,/\,Exclude of a member from a secured group}
\label{sec:concept-update-keys}
Withdrawal of a member from a group can be initiated by the participant herself or be determined by the GC as exclusion.
In any case, the presently known $GTEK_{Current}$ and $GKEK_{Current}$ cannot be used, as the expelled participant is in possession of them.
To reduce the effort, CAKE uses a reduced CRT system and a ternary tree structure, which is managed by the GC.
%In expectation of repeated withdrawals, the following process provides performance advantages instead of a re-initialization of the group.

\begin{figure*}[]
	\centering
	\captionsetup{justification=centering}
	\includegraphics[width=1.00\textwidth]{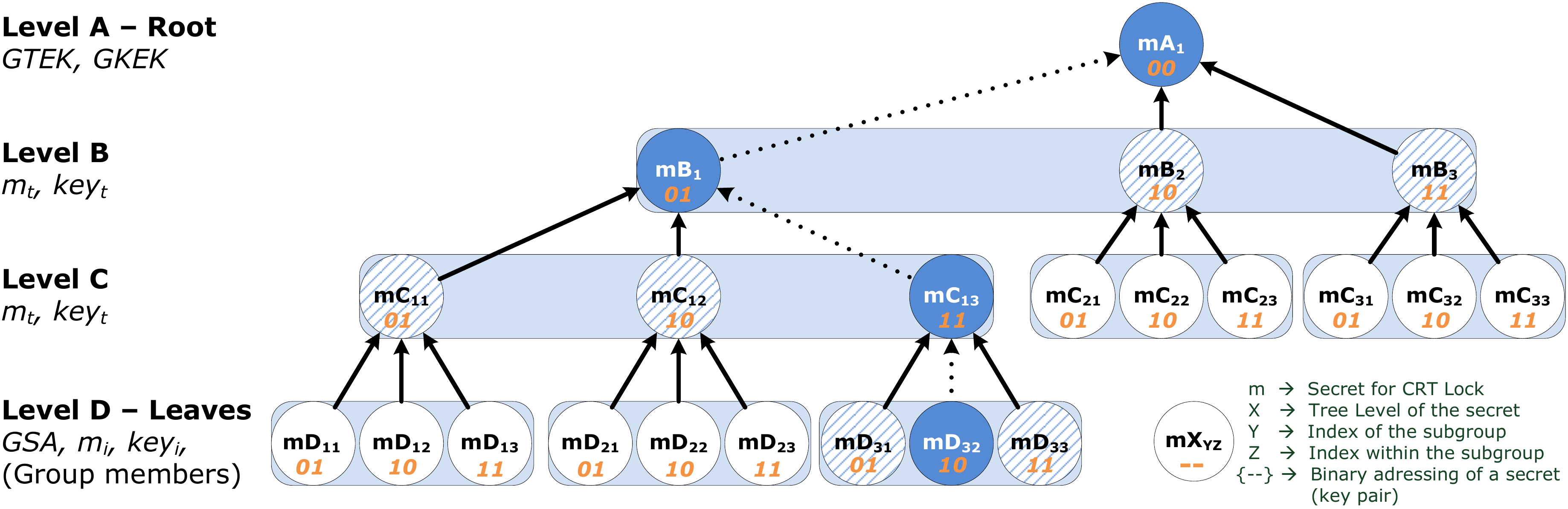}
	%\vspace*{-0.1cm}
	\caption{Ternary tree structure to manage the keys and to reduce the calculation effort by withdrawal.}
	\label{fig:tenerytree}
	%\vspace*{-0.1cm}
\end{figure*}

Figure~\ref{fig:tenerytree} illustrates CAKE's tree structure with level~A (the root) representing the GKEK and GTEK.
Every node represents a pair of keys ($m_t$ and $key_t$) known by the underlying participants.
The actual group members with their personal secrets $m_i$ and $key_i$ are mapped to the leaf nodes of the tree.
The designation mX of a node defines a specific $m_i$ for the CRT system.
%
%The positions in the tree are indexed with binary adresses. The orange number pair is shown specific to prevent problems in case of tree extensions.

All pair of keys on the path from the root to the participant must be known by the participant.
The tree structure enables efficiency, but its creation can be deferred and only be initialized if necessary.
This allows the tree being set up and distributed during a period of low network load.
Considering the state of the art, nearly any tree-based scheme ignores this issue and excludes the costs for the tree setup in the evaluation.

Due to their flat structure, trees with more than two subnodes are better suited for larger groups than binary trees.
In most scenarios (rarely more than 60 participants and hard to imagine more than 300~\cite{economist}), the ternary tree structure is ideal with regard to the size of the tree.

\begin{figure}[t]
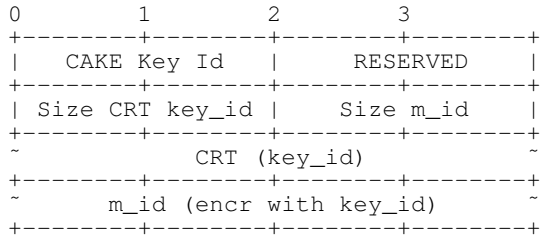

		\centering
		\fontsize{9pt}{9pt}{\selectfont}
		%[fontsize=\tiny]
		\begin{BVerbatim}                                            
0        1        2        3        
+--------+--------+--------+--------+
|   CAKE Key Id   |     RESERVED    |
+--------+--------+--------+--------+
| Size CRT key_id |    Size m_id    |
+--------+--------+--------+--------+
~            CRT (key_id)           ~
+--------+--------+--------+--------+
~      m_id (encr with key_id)      ~
+--------+--------+--------+--------+
		\end{BVerbatim}
		\caption{Exemplary CAKE Keys Substructure}
		\label{fig:cake-keys-substructure}
\end{figure}

\begin{figure}[t]
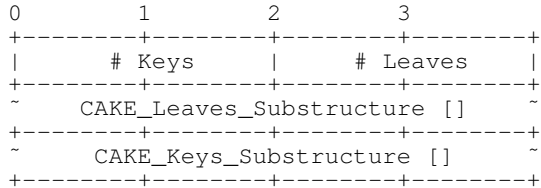

		\centering
		\fontsize{9pt}{9pt}{\selectfont}
		%[fontsize=\tiny]
		\begin{BVerbatim}                            
0        1        2        3        
+--------+--------+--------+--------+
|      # Keys     |     # Leaves    |
+--------+--------+--------+--------+
~    CAKE_Leaves_Substructure []    ~
+--------+--------+--------+--------+
~     CAKE_Keys_Substructure []     ~
+--------+--------+--------+--------+
		\end{BVerbatim}
		\caption{Exemplary CAKE Leave Array}
		\label{fig:cake-leave-array}
\end{figure}

%\begin{figure}[t]
%\begin{subfigure}[t]{0.5\columnwidth}
%\centering
%\fontsize{5.4pt}{8pt}{\selectfont}
%%[fontsize=\tiny]
%\begin{BVerbatim}                                            
%  0        1        2        3        
% +--------+--------+--------+--------+
% |   CAKE Key Id   |     RESERVED    |
% +--------+--------+--------+--------+
% | Size CRT key_id |    Size m_id    |
% +--------+--------+--------+--------+
% ~            CRT (key_id)           ~
% +--------+--------+--------+--------+
% ~      m_id (encr with key_id)      ~
% +--------+--------+--------+--------+
%\end{BVerbatim}
%\caption{CAKE Keys Substructure}
%\label{fig:cake-keys-substructure}
%\end{subfigure}%
%\begin{subfigure}[t]{0.5\columnwidth}
%\centering
%\fontsize{5.4pt}{8pt}{\selectfont}
%%[fontsize=\tiny]
%\begin{BVerbatim}                            
% 0        1        2        3        
%+--------+--------+--------+--------+
%|      # Keys     |     # Leaves    |
%+--------+--------+--------+--------+
%~    CAKE_Leaves_Substructure []    ~
%+--------+--------+--------+--------+
%~     CAKE_Keys_Substructure []     ~
%+--------+--------+--------+--------+
%\end{BVerbatim}
%\caption{CAKE Leave Array}
%\label{fig:cake-leave-array}
%\end{subfigure}%
%\vspace{-1em}
%\caption{Exemplary CAKE Structures for G-IKEv2}%
%\label{fig:cake-structures}%
%\end{figure}
\subsection{Tree Management and Key Addressing}\label{sec:concept-tree-mgmt}
%\peter
In order to differentiate between different keys, an efficient addressing scheme is mandatory.
This applies to every key, not just the keys in the tree.
The ID of every key pair is defined as an $8 x 2$\,bits address.
This address space allows a maximum tree depth of 8 and 2,187 group members in a group with 3,280 key pairs in the tree.
The IDs are starting from \texttt{00} at the root key pair on the top.
Every parent has the children \texttt{01}, \texttt{10} and \texttt{11}.
The unused bits are padded with \texttt{00}.
This allows a unique identification of the position of every key within the tree.

%Additionally, keys required for re-keying actions can be derived easily.
Temporary keys can be derived easily like the keys for re-keying actions.
These key pairs not yet included in the tree structure obtain the ID starting with \texttt{11}.
%Nodes having key pairs not yet included in the tree receive an ID \textit{not} starting with \texttt{00}, but e.\,g. with \texttt{11}.
After a client is authenticated with the GC, it has its own secret $m_i$ and $key_i$ (see Section~\ref{sec:concept-join}), distributed with an address within or outside the tree.
%After being authenticated with the GC, every participant has his own secret $m_i$ and $key_i$ (see Section~\ref{sec:concept-join}), distributed with an address within or outside the tree.
This support the re-balancing if the tree if necessary by the GC.
%This can be implemented by the GC, which may decide to construct and re-balance the tree only if necessary.

In order to take full advantage, the CAKE protocol for the distribution of keys requires the implementation of the following messages:
%Due to the addressing scheme, the protocol for the distribution of keys and performing re-keying requires the implementation of the following actions:
\begin{enumerate*}[label=\Roman*.)]
	\item Downloading key pairs on the path to the root from the GC - \textit{CAKE\_Download\_Array}
	\item Re-Addressing of keys - \textit{CAKE\_Update\_Array}
	\item Receiving updates of key pairs - \textit{CAKE\_Readress\_Array}
	\item Re-Keying upon removal of group members - \textit{CAKE\_Leave\_Array}
\end{enumerate*}

%With these four CAKE Download Types and three substructures holding the array elements are defined.
This is similar to the payloads already defined in G-IKEv2 for \textit{LKH Download Type} (see \mbox{Section 4.8.3} in~\cite{yeung-g-ikev2}).
According to this, these four CAKE Download Types and three substructures are defined as array elements. This support the compatibility.
As shown above, any message can be distributed securely as broadcast.
Furthermore, the information of all download types are structured as an array to differentiate between the elements. 
Each elements of the arrays is indexed with the identity of the key pair in the tree.
Using the unique tree IDs, the nodes can detect if they are affected by the action or not by comparing address prefixes.

The \textit{CAKE\_Download\_Array} can be embedded in GSA\_AUTH or GSA\_REKEY message.
It holds multiple keys, transported within a Key-Substructure (see Figure~\ref{fig:cake-keys-substructure}).
Therefor, a CRT for $m_i$ and $key_i$ is built using the key pairs of the child nodes in the tree.
The substructure includes the address of the key pair and the CRTs themselves.
For efficiency, the tree has to be distributed bottom up as other tree based group key management schemes.
If a client is in possession of one key pair, then he is able to solve the Lock MX to obtain the information.

For the re-addressing of the keys, we use the \textit{CAKE\_Update\_Array} embedded in a GSA\_REKEY message.
Due to the message type, the receiver knows, that there have been distributed keys before.
Thus, these keys need to be updated in the client's CAKE\_SA.

To change the ID of a key pair, the \textit{CAKE\_Readress\_Array} is used.
With this message, a key pair obtains a new position in the tree.
Furthermore, it can be used to include nodes to the tree or to re-arrange subtrees (e.\,g. when the tree is re-balanced).
The readdress information is holding in tuples $(Id_{old}, Id_{new})$.
If a key on his path to the root receives re-addressing, it needs to update all keys on its path further down the tree.

When a node leaves or has to be excluded, a new root key pair has to be created.
It is used to encrypt the GTEK which is embedded in a KD Payload. % (see Section~\ref{cake:rekeying}).
This information is communicated with the \textit{CAKE\_Leave\_Array} message.
A leaving node is in possession of all key pairs on the path to the root (see Figure~\ref{fig:tenerytree}: node mD32, dark, binary address: 00-01-11-10).
All these marked key pairs can not be used for further security operations.
Instead, all mX located next to a marked node on the same level are used (see hatched nodes in Figure~\ref{fig:tenerytree}).
This is done for the entire tree along the path to all key pairs, the leaving client is in possession of.
The updated key pairs are embedded in a Keys-Substructure (see Figure~\ref{fig:cake-keys-substructure}) which in turn is embedded in the \textit{CAKE\_LEAVE\_ARRAY} (see Figure~\ref{fig:cake-leave-array}).
%Besides the Key Array, it also holds an array of leaving nodes which size is at least one.
The operation allows excluding multiple nodes with only one message.

Please note that the GC may choose to distribute only the root key pair with a short and tiny message.
The other keys on the path are updated later with \textit{CAKE\_Update\_Array} instead of one large \textit{CAKE\_Leave\_Array}.
%\begin{enumerate*}[label=\Roman*.)]
%	\item to distribute all updated key pairs in one \textit{CAKE\_Leave\_Array}, or 
%	\item to distribute only the root key pair and update the keys on the path later with \textit{CAKE\_UPDATE\_ARRAY}.
%\end{enumerate*}

\subsection{Merging and splitting groups}
%\peter
To merge two or more existing groups, it requires a renewal of several keys.
The new common $GTEK_{new}$ will be spread based on the currently used individual GKEKs of the merging groups.
It is mandatory to send one message per group to create a merged group.

A group split is done by re-addressing the sub-keys within the key tree and building Lock MX in the amount of divivded subgroups including the new keys.
The number of messages may be as small as one, but depends heavily on the previous tree structure.
As every leave operation, it is coherent with a high effort and will be analyzed in detail in further studies.

Please note that the usage of the \textit{Delete Payload} message as specified in~\cite{yeung-g-ikev2} is not resistant to malicious attacks of internal group members.
In the cryptographic community, this problem is referred to as Post Compromise Security, which is an unresolved unresolved problem.
So CAKE avoids the usage of this message and the process has to be authorised through the group controller.

%\begin{landscape}

\newcommand{\tablebulletpoint}{\quad $\lhookrightarrow$\,}
\newcommand{\tbp}{\mbox{~~~~}}
\newenvironment{tabitemize}{
	\begin{minipage}[t]{\linewidth}%
	\begin{itemize}[itemsep=-2pt,topsep=2pt,leftmargin=5pt,label=\scriptsize{\textbullet}]
}{
	\end{itemize}%
	\end{minipage}
}

\newcommand{\tabitemOne}[1]{
	\begin{tabitemize}
		\item #1
	\end{tabitemize}
}

\newcommand{\tabitemTwo}[2]{
	\begin{tabitemize}
		\item #1
		\item #2
	\end{tabitemize}
}

\begin{table*}[htb]%
\centering
\caption{Comparison of CAKE with LKH and traditional GKMP (represented by G-IKEv2), with special regards on cryptographic overhead. The keys are defined for AES with 16 Byte keys.}
\label{tab:eval-comparison}
\setlength{\tabcolsep}{0.18em} % for the horizontal padding
\renewcommand{\arraystretch}{1.0}% for the vertical padding
\begin{threeparttable}[htb]
\begin{tabular}{l||L{1.5cm}|L{2cm}|L{3.6cm}|L{2.3cm}|L{3.8cm}}
						&	\textbf{Register/Join} &\textbf{Mass Join}\tnote{2} & \textbf{Key Download/Update}\tnote{2} &	\textbf{Tree Operation} & \textbf{Leave}\tnote{2}	\\\hline
Networking
						&	\multicolumn{5}{c}{KD Payload\tnote{1} : 12 (in Bytes)}		\\\hline
						
\tbp GKMP			
						&	KEK: 16\newline 								%Join
							TEK: 16\newline	
							Key:  16
						&	$p$ Messages with:\newline
							KEK: 16\newline 								%Mass Join
							TEK: 16
						& $n$ Messages with:\newline
							KEK: 16\newline 								%Re-Key
							TEK: 16
						&	\textit{undefined}										%Tree Operation		
						&	$n-1$ Messages with:\newline						%Leave
							KEK: 16\newline 								
							TEK: 16\\\hline    
						
\tbp LKH	
						&	KEK: 16\newline 								
							TEK: 16\newline
							Key:  16		
						&	$p$ Messages with:\newline
							KEK: 16\newline 								%Mass Join
							TEK: 16									
						&	1 Message with\tnote{3} :\newline
							Hdr:	$(4 + (n-1)*12)$\newline 
							Keys: 	$(n-1)*16$
						&	\textit{same as Key Download}
						&	1 Message with\tnote{3} :\newline
							Hdr: $4 + log_2(n) * 8 + \sum_{i=1}^{log_2(n)-1}i*8$\newline
							Keys: 	$\sum_{i=1}^{log_2(n)-1}i*16$
						\\\hline
\tbp CAKE			
						&		KEK: 16\newline
								TEK: 16\newline
								Prime: 17\newline
								Key:  16				
						&   1 Message with:\newline
								Hdr: 16\newline
								TEK: 16\newline
								KEK\tnote{5} :$|CRT(p)|$							
						&		1 Message with\tnote{3} :\newline
								Hdr:	$4 + n*12$\newline 
								Keys\tnote{5} : $({log_3(n)-1})*|CRT(3)|$\newline
								Primes:	$({log_3(n)-1})*3*17$
						&		1 Message with\tnote{3} :\newline
								Hdr: $(4+8)$\newline
								Key\tnote{5} : $|CRT(3)|$\newline
								Primes: $3*17$
						&		1 Message with:\newline																		
								Hdr: $12 + log_3(n^2) * 8$\newline
								TEK: $16$ \newline
								KEK\tnote{5}: $|CRT(log_3(n^2))|$
						\\\hline\hline
Computation	& & & & \\\hline
\tbp GKMP\tnote{4,7}		
						&	GC: $O_K(1)$ \newline
						  Cl: $O_K(1)$
						& GC: $O_K(p)$\newline
							Cl: $O_K(1)$
						&	GC: $O_K(n)$ \newline
						  Cl: $O_K(1)$
						&	\textit{undefined}											
						&	GC: $O_K(n-1)$\newline
						  Cl: $O_K(1)$
						\\\hline
\tbp LKH\tnote{4,7}			
						&	\textit{see GKMP}
						& GC: $O_K(p)$\newline
							Cl: $O_K(1)$
						&	GC: $O_K(2^{log_2(n)+1})$\newline
						  Cl: $O_K(log_2(n)+1)$
						&	\textit{same as Key Download}																					
						&	GC: $O_K((log_2(n)+log_2(n-1)))$\newline
						  Cl: $O_K(log_2(n))$ 												
						\\\hline
\tbp CAKE\tnote{6,7}	
						&	\textit{see GKMP}
						& GC: $O_L(p)$\newline
							GC: $O_K(1)$\newline
							Cl: $O_L(p)$\newline
							Cl: $O_K(1)$
						& GC: $O_L(3*\frac{n-1}{2})$\newline
							GC: $O_K(1)$\newline
							Cl: $O_L(3*log_3(n))$\newline
							Cl: $O_K(1)$
						& GC: $O_L(3)$\newline
							GC: $O_K(1)$\newline
							Cl: $O_L(3)$\newline
							Cl: $O_K(1)$
						& GC: $O_L(log_3(n^2))$\newline
							GC: $O_K(1)$\newline
							Cl: $O_L(log_3(n^2))$\newline
							Cl: $O_K(1)$
						\\\hline

\end{tabular}
\centering
\par\medskip
\begin{tablenotes}
	\item[1] Required for every Key distributed with G-IKEv2
	\item[2] $n$ being Group Members, $p$ number of members joining or leaving
	\item[3] KEK and TEK is carried as in GKMP
	\item[4] GC performs \textit{encrypt} and Client performs \textit{decrypt}
	\item[5] $|CRT(i)|$: size of CRT with $i$ elements in Bytes
	\item[6] $O_{L}$: Complexity of creating/solving Lock MX. 
	\item[7] $O_K$: Complexity of encryption/decryption of keys.
\end{tablenotes}
\end{threeparttable}
\vspace*{-0.2cm}
\end{table*}

%\end{landscpae}

\section{Evaluation}\label{sec:evaluation}
Having a sound concept at hand, this section evaluates CAKE under the following three aspects:
Firstly, a (theoretical) comparison of the computational complexity as well as networking load of CAKE, LKH and traditional GKMPs is carried out.
Secondly, an implementation of CAKE for RIOT~OS proofs both its lightweight nature and its applicability in constrained scenarios.
Given the result, the section will close by evaluating CAKE against the requirements as stated in Section~\ref{sec:scenario}.

\subsection{Comparison with LKH and GKMP}
Existing concepts for managing group keys are traditional GKMP systems (represented by G-IKEv2) and LKH. 
Table~\ref{tab:eval-comparison} contains the comparison of CAKE with these two concepts regarding the networking and computation overhead. 
On the one side, it is indisputable that as no client leaves the group the traditional GKMP approach performs optimal. 
On the other side, LKH and CAKE perform far better in terms of quantity of messages and computations when \textit{Forward Secrecy} is required at the moment clients leave. 
This is a major benefit of these two concepts.

Although CAKE requires a pair of keys ($m_i$ and $key_i$) to be sent when distributing the tree, it can outperform the LKH mechanism introduced in G-IKEv2.
The amount of key headers is equal in both systems.
Unfortunately, LKH tree entries need to be transported multiple times decreasing its efficiency. % (for better insights to LKH in G-IKEv2, we recommend Appendix~A of~\cite{yeung-g-ikev2}).
Using a CRT system, CAKE offers the distribution of keys using a single message.
Beside this, the message can be send to a later point of time, when network load is low.
Although, the size of the resulting Lock~MX increases linearly (see Table~\ref{tab:evaluation-crt}), it still decreases the necessary protocol information heavily.
%This was one requirement (see Section~\ref{sec:scenario}) and design goal.

CAKE also reduces the demand for computational power on the client-side. 
Instead of carrying out multiple decryption operations (as for example LKH would do), the client has to perform one single modulo and one decrypt operation only.
Nonetheless, this comes at the price of storing more cryptographic material $(m_i, key_i)$ compared to LKH where only $key_i$ has to be stored for every node in the tree.

\subsubsection{Network overhead of LKH}
As LKH operates very similar as CAKE, the following will comparing both overheads.
LKH uses a binary tree and its G-IKEv2 extension can currently handle a maximum number of 65,536 participants.
For comparison we assume a tree with 11 levels resulting in a maximum of 2,048 participants.
Removing one client from the key will therefore result in 10\,keys having to be changed and distributed in the network (for better insights to LKH in G-IKEv2 we recommend Appendix~A of~\cite{yeung-g-ikev2}).
This would result in 10 LKH\_UPDATE\_ARRAYs carrying a total of \mbox{$\sum_{i=1}^{10}16*i+8*i+8 = 1,400$\,Bytes} (16, being the Key size, 8 being the LKH Keys header and 8 being the LKH\_UPDATE\_ARRAY header).
Please note that in the current version of the G-IKEv2 extension for LKH, many keys are transported multiple times which heavily decreases efficiency. 
With some optimizations, the LKH Key Download could be decreased to 464\,Bytes and, thus, being slightly more efficient than CAKE in terms of networking.
However, this slightly better efficiency comes with the cost of higher computation overhead on the client, as in the worst case, 10 keys have to be decrypted individually.
Additionally, changing the tree can currently not benefit from an address scheme as proposed in CAKE, making this operation more expensive in terms of networking and computation.

\subsubsection{Unoptimized re-key networking overhead}
Using the G-IKEv2 protocol without any optimizations (such as CAKE or LKH) would result in number of participants messages including 80\,Bytes overhead for KEK and TEK.
Assuming the current maximum of 2,187 participants, this would result in 2,187 messages.
The IETF draft for G-IKEv2~\cite{yeung-g-ikev2} defines an additional GSA\_INBAND\_REKEY message for such tasks.
Even without optimization, the new keys could be carried out in a single broadcast message, encrypting the KEK with every privately shared secret between client and server.
The GSA\_REKEY message would carry one KEK Key Download Types (40\,Bytes) for every participant and one TEK Key Download Type (40\,Bytes).
For 2,187 participants, this would result in a $2,187*40 + 40=87,520$\,Bytes overhead.

\subsection{Performance on constrained hardware}
RIOT OS~\cite{Baccelli.2013} is an open source operating system that supports various hardware.
Its minimal requirement of 1.5 KB main memory illustrates its lightweight nature and is one of the reasons we implemented CAKE on RIOT.
Further, necessary cryptographic libraries with importance for embedded systems are available.
The generation of an evaluation environment with realistic conditions is achieved by using IOT-LAB~\cite{Adjih.2015}.
It provides a huge amount of wireless nodes with minimal capabilities regarding CPU and memory.
The IOT-LAB M3 board comes with a 72~MHz CPU and 64~KB SRAM and is used for the GC and all clients.
The evaluation focuses on the group management and the associated key distribution processes.
%For a more detailed consideration of the initial key exchange with G-IKEv2 we refer to~\cite{gentschendFelde.2017}.

\subsubsection{Memory requirements}
For the evaluation we created a homogenous setup with a GC and a group of 14 clients on IOT-LAB M3 nodes.
Regarding the memory requirements, the design principles of RIOT OS need to be considered, as all memory is statically reserved, including the network buffer.
The required memory on the GC is defined through the number of all necessary keys within the ternary tree including the nodes.
For each participant, the GC requires a total of 2,900~Bytes of data being stored, consisting of keys, IP addresses, and memory for CRT calculations and tree operations.
Subsequently, the required memory for the GC is 40,600~Bytes in total, which is covered by the available memory of 64~KB in our evaluation setup.
The memory requirements for participants of the group are lower.
Each client requires only 2,900~Byte per connection to a GC.

\subsubsection{Computational Costs for CRT}
\begin{table}[t]%
\caption{Required time for Lock MX operations with $i$ elements. For comparison, the time to encrypt and decrypt the key hierarchy of LKH with tree depth $i$ is shown. The number of clients is $3^{i}$ for CAKE and $2^{i}$ for LKH.}
\label{tab:evaluation-crt}
\centering
%\begin{tabular}{L{1cm}|R{2cm}|R{2cm}|R{2cm}}
%\begin{threeparttable}[b]
\setlength{\tabcolsep}{0.4em} % for the horizontal padding
\renewcommand{\arraystretch}{0.4}% for the vertical padding
\begin{tabular}{r|R{1.5cm}|R{1cm}|R{1cm}|R{1cm}|R{1cm}}
$i$ 				&	Create\newline Lock MX\newline (\,$\mu$s) 			&	Solve Lock MX\newline (\,$\mu$s) &	Size\newline Lock MX\newline (\,Byte) 	&  LKH Enc\newline (\,$\mu$s)  		& LKH Dec\newline (\,$\mu$s)\\\hline
1						&    280,082					&	   88	& 	 41 					&  125	&  201\\
2						&    572,785					&	  189	& 	 84					&  188	&  302\\
3						&    822,851					&	  275	& 	124					&  250	&  404\\
4						&  1,130,065					&	  374	& 	165 					&  312	&  505\\
5						&  1,377,708					&	  484	& 	206 					&  374	&  607\\
6						&  1,539,600					&	  604	& 	247 					&  437	&  708\\
7						&  1,909,062					&	  750	& 	288 					&  499	&  809\\
8						&  2,231,764					&	  904	& 	328 					&  562	&  911\\
9						&  2,544,507					&	1,072	& 	369 					&  624	&1,013\\
10					&  2,751,188					&	1,243	& 	410 					&  686	&1,114\\
11					&  3,134,233					&	1,433	& 	451 					&  749	&1,215\\
12					&  3,387,458					&	1,632	& 	492 					&  811	&1,316\\
13					&  3,705,136					&	1,858	& 	533 					&  874	&1,418\\
14					&  3,974,770					&	2,081	& 	573 					&  935	&1,520\\
\end{tabular}
%\centering
%\par
%\begin{tablenotes}
	%\item[*] Not a valid tree operation in CAKE.
%\end{tablenotes}
%\end{threeparttable}
\end{table}

As the evaluation focuses on constrained hardware, we concentrate on time measurements for the cryptographic calculations.
The most expensive operation is the IKE SA INIT message, which is caused by the necessary computation of the DH key exchange.
The measurements in our evaluation setup show computation times on the IOT-LAB M3 nodes that are comparable with the times on Arduino Due in~\cite{gentschendFelde.2017}.
Furthermore, most actions in CAKE require only one single GSA\_REKEY message carrying G-IKEv2 payloads (see~\cite{yeung-g-ikev2}), which is beneficial in terms of computation time.

The most interesting new feature of CAKE is the Lock MX creation and solving on the GC and the clients in terms of computational cost.
Table~\ref{tab:evaluation-crt} shows the measurement results for the creation of the Lock MX with different tree depths as well as the time to resolve it on client side.
These results are especially notable regarding a \textit{mass entry} (see Section~\ref{sec:concept-join}), where $i$ represents the number of clients joining simultaneously. 
Along with the number of elements in the CRT, the computation time of new keys increases.
Evidently, this is mainly caused by the higher size of the Lock MX.
On the other side, the clients significantly benefit from the new method to receive keys.
One simple modulo operation is needed on client side resulting in low computation time for solving the Lock MX even at its maximum size of 14 elements.

For comparison, the costs for encrypting and decrypting keys within an LKH tree are shown in Table~\ref{tab:evaluation-crt}. 
It can be seen that even though the AES implementation is highly optimized, solving the Lock~MX scales similar to decrypting the keys within the LKH tree.
However, lowering network load with CAKE comes at the price of computational overhead for creating the Lock~MX, which scales worse than LKH.
Optimizing the Lock~MX implementations will be part of further studies.

\subsection{Fulfillment of requirements}
Resource-constrained environments necessitate functional and non-functional requirements.
The design of CAKE particularly focuses to meet these.
Firstly, the ternary tree enables a reduced number of keys needed to be stored and sent.
Additionally, through CAKE a re-keying is possible with one single message and the per-packet overhead is reduced.
Thus, the requirements \textbf{Low Datarates} and \textbf{No 1-to-n Effect} are fulfilled.
When group changes happen, the evaluation shows that CAKE requires only a minimum of messages to be sent.
For group leave actions the CRT is utilized and the calculation complexity is reduced to only one modulo operation on client side.
So, \textbf{Minimal Delay} and \textbf{Low calculation complexity} are achieved. 
The required \textbf{Minimal amount of key changes and exchanges} is realized by combining the ternary tree and the addressing scheme. 
This allows the minimization of required actions in case of restructuring the tree.
Moreover, the \textbf{Compatibility} requirement is achieved by using the G-IKEv2 protocol, which is currently being standardized.
Any client that is not capable of the newly introduced features may participate in the group by utilizing standard re-keying mechanism, while the CAKE-capable clients can still use the optimized feature set.
Additionally, through the use of G-IKEv2, the \textbf{Security Properties} are met, as they are included in the standardization and therefore well studied.
From this point of view, CAKE is an optimization of key calculation and transport, leaving the security parameters as they are.
Lastly, the \textbf{Minimal Trust} requirement can be accomplished on a per-scenario-basis throughout the various supported authentication mechanisms of G-IKEv2.

Besides the practical applicability and its pros and cons in comparison to especially LKH, reviewing the initial design goals and requirements shows completeness. 
A detailed design explanation and security assessment can be found  in \cite{Hillmann2017}.

%\begin{figure*}[t]
	%\begin{subfigure}[t]{1\columnwidth}
		%\centering
		%\includegraphics[width=\linewidth]{crt-time-measure-server}
		%\caption{create the Lock~MX on server.}
		%\label{fig:evaluation-create-mx-server}
	%\end{subfigure}%
	%\begin{subfigure}[t]{0.95\columnwidth}
		%\centering
		%\includegraphics[width=\linewidth]{crt-time-measure-client}
		%\caption{resolve CRT on client.}
		%\label{fig:evaluation-resolve-mx-client}
	%\end{subfigure}%
%\vspace{-1em}
%\caption{Required times to create and solve the Lock~MX on M3 nodes.}
%\label{fig:evaluation-crt}
%\end{figure*}

%\begin{figure}[bhpt]
%	%{\centering
%	\centering
%	\captionsetup{justification=centering}
%	\includegraphics[width=0.47 \textwidth]{picture/20180206-164901.jpg}
%	\caption{Why secure group communication.}
%	\label{fig:20180206-164901}
%	%}
%	%\vspace*{-0.2cm}
%\end{figure}

\section{Conclusion}\label{sec:conclusion}
%\peter\todo[inline]{ca 0.5 Seiten}
In this paper, we presented the concept of an efficient group-key-management protocol that meets the requirements of resource-efficient procedures in many application scenarios like MANET.
It is based on the combination of the lightweight G-IKEv2~\cite{RFC7296} communication protocol in combination with CAKE~\cite{Hillmann2017} for the key exchange and key management.
CAKE offers the possibility to exchange keys within a group and to react efficiently to dynamic changes of the group with low calculation effort and a low load on the network.
Nevertheless, it enables confidential key distribution and compliance with backward and forward security requirements for mobile computing.
The main objective to reduce the network load to a minimum is achieved at the cost of additional storage space for supplement cryptographic key material.
The CRT-based key hierarchy together with a ternary keys tree structure reduce the data to be transferred especially during group leave operations.
The design of CAKE delegates computational demanding cryptographic operations to the group controller, relieving to potentially less powerful group members.
In today's interconnected world, this middleware technology shows advantageous in the area of secure group communication among highly constrained group members.

In the current state of research, the inefficient addressing scheme will be optimized in the next step.
Apart from that, the networking overhead of the LKH extension in G-IKEv2 can be also further improved.
Thus, the optimization of both in concerning scenarios is subject of future work, which will allow for a more comprehensive and technical comparison of LKH and CAKE.
Beside further improvements of the CAKE prototype, the basic concept and the implementation is evaluated for building and solving the CRT System worth investigating.
Finally, a more detailed analysis of the solution to post-compromise security is of great interest in the case of merging and division of groups.

\section*{Acknowledgment}
We thank the MNM-Team for all the helpful comments and discussions while writing this document, especially Nils gentschen Felde.
We also thank our students Edgar Goetzendorff, B.Sc., and Mehdi Yosofie, B.Sc., for assisting us during the development of the prototype and the helpful discussions.

\bibliography{literature,rfc,lniguide}
\end{document}